# On the possible void decay in free-electron laser sase-fel experiment

J. Marciak – Kozlowska [1], M. A Pelc [2]
M. Kozlowski [3],

[1] Institute of Electron Technology,
Al Lotnikow 32/46, 02-668 Warsaw, Poland
[2] Institute of Physics, Maria Curie-Sklodowska University,
Lublin, Poland
[3] correresponding author:     e-mail; miroslawkozlowski@aster.pl

Abstract
 In this paper the motion of ultra – high energy particles produced in SASE-FEL is investigated.  The critical field which opose the acceleration of the ultra high energy particles is calculated

*We must make some profound alterations to the theoretical idea    of the vacuum. . . . Thus, with the new theory of   electrodynamics we are rather forced to have an    aether"*

**Introduction**

Recently the ultra-high energy lasers proposals are developed [1]. The SASE-FEL project for the first time enables the investigation of the "vacuum decay" processes and emission of ultra-relativistic fermions. In this paper we argue that the results of the SASE-FEL future experiments opens the new field of the investigation of the structure of the spacetime

**The model description**

Bell's theorem is rooted in two assumptions: the objective reality – the reality of the external world, independent of our observations; the second is locality, or no faster than light signaling. Aspect's experiment appear to indicate that one of these two has to go.

In this paper we are going back to relativity as it was before Einstein when people like Lorentz and Poincaré thought that there was an aether – a preferred frame of reference – but that our measuring instruments were distorted by motion in such a way that we could not detect motion through the aether. Now in that way you can imagine that there is a preferred frame of reference and in this preferred frame of reference particles do go faster than light. But then in other, our, frame of reference particles have the speed lower than the light speed.

In this paper we propose the following scenario. Behind the scene – our world of observation something is going which is not allowed to appear on the scenes.

To start with we observe that for electrons, at the vicinity of $E_k \sim mc^2$ the speed of electrons is changed abruptly. With $E_k \sim mc^2$, through Heisenberg inequality the characteristic time can be defined

$$\tau = \frac{\hbar}{mc^2} \tag{1}$$

At that characteristic time the Newton theory (NT) and SR theory starts to give different description of the speeds, viz.,

$$\upsilon_{NT} = \sqrt{\frac{2E_k}{m_k}} \tag{2}$$

and

$$\upsilon_{SR} = c\sqrt{1 - \frac{1}{\left(\frac{E_k}{mc^2}+1\right)^2}} \tag{3}$$

Let us introduce the acceleration $a_m$ which describe the change of speeds in time $\tau$

$$a_m = \frac{\upsilon_{NT} - \upsilon_{SR}}{\tau} \tag{4}$$

and force $F$ which opose the acceleration of the particle with mass $m$

$$F = ma_m = \frac{mc}{\tau}\left(\left(\frac{2E_k}{mc^2}\right)^{\frac{1}{2}} - \left(1 - \frac{1}{\left(\frac{E_k}{mc^2}+1\right)^2}\right)^{\frac{1}{2}}\right)$$

$$= \frac{m^2 c^3}{\hbar}\left(\left(\frac{2E_k}{mc^2}\right)^{\frac{1}{2}} - \left(1 - \frac{1}{\left(\frac{E_k}{mc^2}+1\right)^2}\right)^{\frac{1}{2}}\right) \tag{5}$$

In formula (5) we introduce the field $E_s$

$$E_s = \frac{m^2 c^3}{\hbar e} \tag{6}$$

and we obtain:

$$F = E_s e = \left(\left(\frac{2E_k}{mc^2}\right)^{\frac{1}{2}} - \left(1 - \frac{1}{\left(\frac{E_k}{mc^2}+1\right)^2}\right)^{\frac{1}{2}}\right) \tag{7}$$

It is interesting to observe that the field $E_s$ is the same as the Schwinger field strengths [2]. J. Schwinger demonstrated that in the background of a static electric field, the QED vacuum is broken and decayed with spontaneous emission of $e^+e^-$ pairs.

In the following we define for electrons the energy $L$

$$L = eF_s r_e = \alpha \frac{(m_e c^2)^4}{e^6} \tag{8}$$

In the formula (8) $r_e$ is the classic electron radius

$$r_e = \frac{e^2}{m_e c^2} \tag{9}$$

and $\alpha$ is the fine structure constant. Having the energy $L$ and volume $r_e^3$ where the energy is concentrated we define the bulk modulus for the medium which oppose the motion of electron

$$B = \frac{L}{r_e^3} = \alpha \frac{(m_e c^2)^4}{e^6} \tag{10}$$

and hypotetic sound velocity in the medium which opose the acceleration of the electrons

$$\upsilon_{sound} = \left(\frac{B}{\rho}\right)^{\frac{1}{2}} = 10^{18} c \qquad (11)$$

where $\rho = \dfrac{m}{r_e^3}$.

**Conclusion**

In this paper we argue the existence of the medium with the following properties: bulk modulus ~ $10^{37}$ MeV/cm³, sound velocity $10^{18}$ c.

**References**

[1] TESLA , SASE-FEL Technical Design Project
[3] J. Schwinger, Phys. Rev. *82* (1951) 664